\documentclass[10pt,twocolumn,twoside]{IEEEtran}
\usepackage{graphicx,amssymb,amsthm,graphicx,subcaption}
\ifCLASSINFOpdf
\else
\fi
\usepackage[cmex10]{amsmath}
\begin{document}
%
\title{Covariance-Domain Dictionary Learning for Overcomplete EEG Source Identification}
%
%
%

\author{Ozgur~Balkan*,~\IEEEmembership{Student Member,~IEEE,}
        Kenneth~Kreutz-Delgado,~\IEEEmembership{Fellow,~IEEE,}
        and~Scott~Makeig

\thanks{This work was supported by The Swartz Foundation (Old Field, NY). O. Balkan and K. Kreutz-Delgado are with the Department of Electrical and Computer Engineering, University of California San Diego, La Jolla CA, 92093 USA e-mail: (obalkan@ucsd.edu, kreutz@eng.ucsd.edu). S. Makeig is with the UCSD Swartz Center for Computational Neuroscience email: (smakeig@ucsd.edu).}}
\maketitle

\begin{abstract}
We propose an algorithm targeting the identification of more sources than channels for electroencephalography (EEG). Our overcomplete source identification algorithm, Cov-DL, leverages dictionary learning methods applied in the covariance-domain. Assuming that EEG sources are uncorrelated within moving time-windows and the scalp mixing is linear, the forward problem can be transferred to the covariance domain which has higher dimensionality than the original EEG channel domain. This allows for learning the overcomplete mixing matrix that generates the scalp EEG even when there may be more sources than sensors active at any time segment, i.e. when there are non-sparse sources. This is contrary to straight-forward dictionary learning methods that are based on the assumption of sparsity, which is not a satisfied condition in the case of low-density EEG systems. We present two different learning strategies for Cov-DL, determined by the size of the target mixing matrix. We demonstrate that Cov-DL outperforms existing overcomplete ICA algorithms under various scenarios of EEG simulations and real EEG experiments.  
\end{abstract}


%
\IEEEpeerreviewmaketitle

\section{Introduction}
%
%
%
%
\IEEEPARstart{A}{s} a non-invasive brain imaging modality, electroencephalography (EEG) provides high temporal resolution, applicability in mobile settings, and direct measurement of electrical brain activity as opposed to other brain imaging modalities such as BOLD activity measured in fMRI. However, a major issue in EEG signal processing is that signals measured on the scalp surface do not each index a single localized cortical source of brain activity. Because of the broad point spread function of generated potentials in the brain, EEG data collected on scalp channels is a mixture of simultaneously active brain sources distributed over many different brain areas. In addition, non-brain sources such as eye and muscle movements contribute to the mixing process as well, which makes direct channel-level EEG analysis problematic. For accurate brain activity monitoring, individual sources involved in the mixture have to be identified and extracted from scalp channel data. 

Because of the fact that volume conduction and mixing at the sensors is linear \cite{nunez1997eeg}, EEG mixing can be formulated as follows
\begin{align}
\mathbf{Y} = \mathbf{AX}
\end{align}
where $\mathbf{Y}\in\mathbb{R}^{M\times N_{d}}$ is the matrix containing collected EEG data at $M$ sensors for $N_{d}$ data points. The matrix $\mathbf{A}\in\mathbb{R}^{M\times N}$ is the unknown mixing matrix, and $\mathbf{X}\in\mathbb{R}^{N\times N_{d}}$ contains the activations of $N$ sources. The i-th column of $\mathbf{A}$, denoted as $\mathbf{a_i}$, represents the relative projection weights of the $i$-th source to each channel. The so-called EEG inverse problem is to identify both $\mathbf{A}$ and $\mathbf{X}$, given sensor data $\mathbf{Y}$ \cite{makeig1996independent}. Learning the columns of $\mathbf{A}$, namely the scalp maps, can further enable source localization in the cortex through methods such as DIPFIT \cite{oostenveld2010fieldtrip} or sLORETA \cite{pascual2002standardized}. Identifying the rows of $\mathbf{X}$ can enable the computing of time-series measures such as event-related potentials (ERPs), event-related spectral perturbation (ERSPs), and spectral components \cite{makeig2004mining}. 

A commonly applied method to solve the EEG inverse problem has been to use independent component analysis (ICA) \cite{makeig1996independent,makeig1997blind}. Assuming statistical independence between source activities, ICA can separate the scalp mixture into underlying source time-series $\mathbf{X}$ and identify the mixing matrix $\mathbf{A}$. It was shown in \cite{jasonDipolar} that ICA methods are well suited for solving the EEG inverse problem since independence among sources was found to be positively correlated with the number of brain sources that can be extracted from data. ICA has been extensively applied on EEG for artifact rejection and source separation \cite{jung2000removing,jung2001imaging} and has been shown to increase accuracy in brain-computer-interface paradigms \cite{wang2013improving}. However, one major drawback of ICA is that the number of mixed sources is assumed to be less than or equal to the number of sensors ($N\leq M$). This assumption undermines the reliability and utility of ICA, especially in low-density EEG systems ($<32$ number of channels). 

There are multiple reasons why an EEG source identification algorithm should be able to handle more sources than sensors. A main motivation is to increase the capabilities of EEG systems to handle large number of artifacts. Depending on the experiment settings and the length of recording, the number of distinct artifact sources could possibly outnumber the brain sources or even exceed the number of channels. In those cases, ICA solution matrix is occupied by artifact sources and only a few brain sources can be extracted from data, which limits further analysis of brain activity. Even in ideal conditions, i.e, when there are no artifacts, higher resolution is desired to better capture true brain dynamics, taking into account the possibility of more than $M$ sources being simultaneously active and/or changing brain source locations throughout the experiment. 

It is also desirable to enhance the capabilities of low-density EEG devices that are becoming increasingly popular due to their relative low-cost and ease of use. Low-density EEG allows for a wide range of applications by facilitating EEG recording of mobile and possibly long duration experiments. However, because they are targeted for low-cost research and consumer markets, these systems usually contain about 8-19 channels for which the results of traditional ICA results would be insufficient for reliable brain source monitoring. Extracting more sources than channels may benefit low-cost clinical research and improve consumer-oriented BCI applications.

Here, we propose a covariance-domain dictionary learning algorithm, Cov-DL, that can identify more sources than number of channels for the EEG inverse problem. We note that our algorithm does not learn the explicit source time-series activity $\mathbf{X}$ but learns the overcomplete mixing matrix $\mathbf{A}$ (projection of sources to scalp sensors) and the power of individual sources in a given data segment. In this sense, our algorithm is categorically placed between blind source identification and source separation methods.

\section{Related Work}
An important family of blind source identification methods is comprised of cumulant-based algorithms that incorporate second order (SOBI) \cite{belouchrani1997blind} or fourth order statistics (FOOBI) \cite{de2007fourth}. In non-EEG settings, it was shown that FOOBI can identify a number of sources that are roughly quadratic in the number of sensors \cite{de2007fourth}. However, multiple studies \cite{jasonDipolar,albera2012ica} showed that cumulant-based methods perform relatively poorly in EEG source separation tasks compared to maximum likelihood based methods such as Infomax \cite{infomax}. Among all methods, AMICA, an EM-based maximum likelihood ICA framework with flexible source densities, \cite{jasonAMICA}, performed best in terms of extracting the most number of plausible brain sources while providing the highest independence among sources \cite{jasonDipolar}. 

An extension of traditional ICA for the overcomplete case is provided by the ICA mixture model \cite{lee2000ica, jasonAMICA}. This approach learns $N_{\text{model}}$ mixing matrices, $\mathbf{A}_i \in \mathbb{R}^{M\times M}$, instead of learning one overcomplete mixing matrix $\mathbf{A}$, in order to provide tractable computation. An adaptation of this method with AMICA, Multiple Model AMICA, was shown to be successful in identifying more sources than electrodes in some non-stationary EEG paradigms \cite{jasonAMICA}. However, the mixture model has some drawbacks; because it assumes that at most $M$ sources are active at any given time and there are only a few disjoint sets of simultaneously active sources ($N_{\text{model}}$). This is problematic especially when $M$ is low. An ideal algorithm should be able to handle cases where any of $N\choose k$ sources, $1\leq k\leq N$, can be jointly active. Our algorithm targets this case. 

Another set of overcomplete ICA algorithms \cite{amari1999natural, le2011ica} model the source estimates as $\mathbf{\hat{X} = WY}$, where $\mathbf{W}\in\mathbb{R}^{N\times M}$ is a tall unmixing matrix with full column rank. These algorithms optimize $\mathbf{W}$ and return the mixing matrix as $\mathbf{A = W^T}$. One of the recent algorithms of this type is RICA \cite{le2011ica}, an efficient method used for unsupervised feature learning in neural networks.  We have found that in the complete mixing matrix case ($M = N$), RICA gives almost identical results with Infomax on EEG data. In this paper, we are considering the overcomplete setting for RICA and multiple model AMICA for comparison with our overcomplete method Cov-DL. 

Dictionary learning-based sparse coding algorithms are closely related to overcomplete ICA methods. In the dictionary learning framework, the inverse problem is formulated as the following optimization problem,
\begin{align}
\min_{A,X} \frac{1}{2}\sum_{t = 1}^{N_d} \|\mathbf{Y - AX}\|_{F}^2 + \lambda \sum_{t = 1}^{N_d}g(\mathbf{x_t})
\label{dictLearn}
\end{align}
where $g(\cdot)$ is a function that promotes sparsity of the source vector $\mathbf{x_t}$ at time index $t$ and $\lambda$ is the regularization parameter controlling the sparsity of the sources. Optimization is generally performed on $\mathbf{A}$ and $\mathbf{X}$ iteratively, namely learning $\mathbf{X}$ while keeping $\mathbf{A}$ fixed, and vice versa \cite{aharon2006img,kreutz2003dictionary}. Given a fixed dictionary $\mathbf{\hat{A}}$, the sources $\mathbf{\hat{X}}$ are learned by solving the following optimization  
\begin{align}
\min_{X} \frac{1}{2}\sum_{t = 1}^{N_d} \|\mathbf{Y - \hat{A}X}\|_{F}^2 + \lambda \sum_{t = 1}^{N_d}g(\mathbf{x_t})
\end{align}
The true dictionary can be recovered if the sources $\mathbf{x_t}$ are sparse ($k_{t}<M$), where $k_t$ is the number of active sources at time $t$. The accuracy of recovery is strongly dependent on the level of sparsity as higher accuracy is achieved if $k\ll M$. It was shown for various dictionary learning algorithms that the performance significantly drops as $k$ approaches $M$ \cite{parker2013bilinear}. Indeed, when $k\geq M$, any full-row rank dictionary can provide a source decomposition with sparsity $k$ and zero representation error $\mathbf{\|\mathbf{Y - AX}\|_{F}^2}$ for \eqref{dictLearn}, thus the true mixing matrix becomes unrecoverable. In the case of EEG, this allows at most k = $\mathcal{O}(M)$ EEG sources to be simultaneously active which limits direct applicability of dictionary learning to low-density EEG systems. 

\begin{figure*}[!t]
\centering
\includegraphics[width=0.95\linewidth]{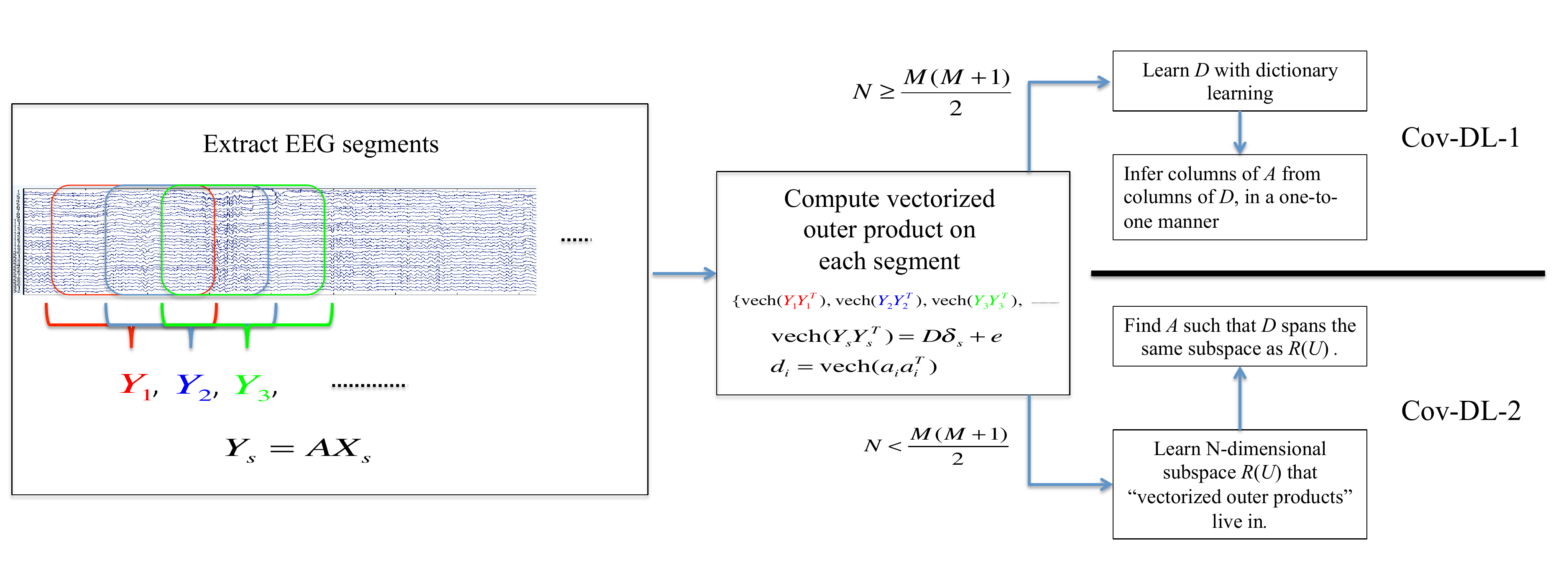}
\caption{The summary of two different strategies of Cov-DL for overcomplete EEG source identification. Cov-DL-1 involves a dictionary learning stage requiring the assumption that $k<M(M+1)/2$ sources are active at any given segment. Cov-DL-2 does not require sparsity of sources.}
\label{fig:flowChart}
\end{figure*}

Recently it was shown that given the true dictionary $\mathbf{A}$, and a data segment $\mathbf{Y_s} \in \mathbb{R}^{M\times L_s}$, where $L_s$ is the length of the segment in data frames, M-SBL (multiple measurement Sparse Bayesian Learning) applied directly on $\mathbf{Y_s}$ can identify active sources under the assumption that sources are uncorrelated in the time segment \cite{balkan2014localization}. The number of sources identified in this case is not limited by the number of channels $M$, $1\leq k \leq M(M+1)/2$. This finding is supported by \cite{pal2015pushing}, where LASSO is applied on the covariance matrix off the data segment $\mathbf{Y_s}$ to obtain probability bounds on the identification of active sources. Under the assumption of uncorrelated sources $\mathbf{X_s}$, the sample-covariance matrix $\frac{1}{L_s}\mathbf{X_sX_s^T}$ is assumed to be nearly diagonal (``pseudo-diagonal") and expressible as $\mathbf{\Sigma_{X_s}} = \frac{1}{L_s}\mathbf{X_sX_s^T = \Delta + E}$, where $\mathbf{\Delta}$ is a diagonal matrix composed of diagonal entries of $\Sigma_{X_s}$. Hence in \cite{pal2015pushing}, $\mathbf{Y_s = AX_s}$ is modeled as 
\begin{align}
\mathbf{Y_sY_s^T} &= \mathbf{AX_sX_s^TA^T}\nonumber \\
\mathbf{\Sigma_{Y_s}} &= \mathbf{A\Sigma_{X_s}A^T} \nonumber \\
\mathbf{\Sigma_{Y_s}} &= \mathbf{A\Delta A^T + E} = \mathbf{\sum_{i = 1}^{N} \Delta_{ii} a_ia_i^T + E}.
\label{corrEq}
\end{align}
Since the covariance matrix is symmetric, we can vectorize the lower triangular part of both sides and obtain,
\begin{align}
\text{vech}\left({\Sigma_{Y_s}}\right) &= \mathbf{\sum_{i = 1}^{N} \text{vech}\left(a_i a_i^T\right)\Delta_{ii} + \text{vech}\left(E\right)}\nonumber \\
\text{vech}\left({\Sigma_{Y_s}}\right) &= \mathbf{\sum_{i = 1}^{N} \mathbf{d_i}\Delta_{ii} + \text{vech}\left(E\right)} \nonumber \\
\text{vech}\left({\Sigma_{Y_s}}\right) &= \mathbf{D\delta + \text{vech}\left(E\right)}
\label{sparseDecompWithUncorrelated}
\end{align}
where $\mathbf{D = [d_1, d_2, \hdots , d_N]}$,  $\mathbf{d_i} = \text{vech}\left(\mathbf{a_i a_i^T}\right)$ and $\text{vech}(\cdot)$ is a function that maps a symmetric matrix $S\in \mathbb{R}^{M\times M}$ to its vectorized lower triangular matrix, of size $\frac{M(M+1)}{2}$. Here, we also define the inverse function $\text{vech}^{-1}(\cdot)$, which takes as an input an $\frac{M(M+1)}{2}$ dimesional vector $v$ and outputs a symmetric matrix of size $M\times M$ whose lower triangular matrix consists of entries in $v$. Thus, for any vector $v$, we have $v = \text{vech}\left(\text{vech}^{-1}\left(v\right)\right)$.  

It was shown in \cite{pal2015pushing} that this formulation, together with the correlation constraint \eqref{corrEq} can identify $\mathcal{O}(M^2)$ sources given the true dictionary. We leverage this idea to also learn the dictionary $\mathbf{A}$ from EEG data considering multiple segments from the overall recording. We also note that assumption of uncorrelated sources, albeit being a weaker constraint, is implied by the independence of sources, an assumption which was shown to be successful for EEG source separation \cite{jasonDipolar}.

\section{Covariance-Domain Dictionary Learning (Cov-DL)}
Here, we describe our covariance based dictionary learning algorithm that leverages the assumed uncorrelated nature of EEG sources. We start by segmenting the overall EEG data matrix $\mathbf{Y}\in \mathbb{R}^{M\times N_d}$, sampled with frequency $S_f$, into possibly overlapping segments $\mathbf{Y_s}\in \mathbb{R}^{M\times {t_sS_f}}$ of $t_s$ seconds, where $s$ denotes the index for the corresponding segment. For each segment, the following equation holds under the linear mixture model of EEG,
\begin{align}
\mathbf{Y_s = AX_s}, \forall s
\end{align} 
and thus, $\mathbf{Y_sY_s^T = AX_sX_s^TA^T}$. Then, we calculate the sample data covariance $\mathbf{\Sigma_{Y_s}} = \frac{1}{L_s}\mathbf{Y_sY_s^T}$, for each segment $s$. We have,
\begin{align}
\mathbf{\Sigma_{Y_s}} &= \mathbf{A\Delta_sA^T + E_s} \nonumber \\
\text{vech}\left(\mathbf{\Sigma_{Y_s}}\right) &= \mathbf{\sum_{i = 1}^{N} \Delta_{s_{ii}} \text{vech}\left(a_ia_i^T\right) + \text{vech}\left(E_s\right)}, \nonumber \\
\text{vech}\left(\mathbf{\Sigma_{Y_s}}\right) &= \mathbf{D\delta_{s} + \text{vech}\left(E_s\right), \forall s}.
\label{cov-dl}
\end{align}
where the vector $\delta_s$ contains the diagonal entries of the source sample-covariance matrix $\mathbf{\Sigma_{X_s}} = \frac{1}{L_s}\mathbf{X_sX_s^T}$, and the matrix $\mathbf{D}\in \mathbb{R}^{M(M+1)/2\times N}$ consists of columns $\mathbf{d_i} = \text{vech}\left(\mathbf{a_ia_i^T}\right)$. Note that, for each segment, the left hand side of the equations are obtained from data while $\mathbf{D}$ and $\mathbf{\delta_s}$ are not known. Our goal is to first learn $\mathbf{D}$ and then find the associated matrix $\mathbf{A}$. We propose two different approaches to recover $\mathbf{D}$ and $\mathbf{A}$ which depend on the relation between the target number of total sources $N$ and the number of channels $M$. See Fig. \ref{fig:flowChart}.
\subsection{Overcomplete $\mathbf{D}$ (Cov-DL-1)}
When $N$, the number of total sources to be identified for the whole EEG session, is larger than or equal to $M(M+1)/2$, $\mathbf{D}$ in \eqref{cov-dl} is overcomplete. If we assume that at any given segment $s$, there are less than $M(M+1)/2$ active sources, namely $\mathbf{\delta_s}$ is sparse, then we can learn $\mathbf{D}$ by applying traditional dictionary learning methods on the set of data points \{$\text{vech}\left(\mathbf{\Sigma_{Y_s}}\right), \forall s\}$. Note that, the sparsity constraint imposed here, that is $k<M(M+1)/2$ is much weaker than the traditional sparsity constraint $k<M$ and is not necessarily violated when $k>M$. 

After learning dictionary $\mathbf{D}$, we can find the mixing matrix $\mathbf{A}$ that generated $\mathbf{D}$ through the relation $\mathbf{d_i} = \text{vech}\left(\mathbf{a_ia_i^T}\right)$. For each column of the dictionary we optimize,
\begin{align}
\min_{\mathbf{a_i}} \| \mathbf{d_i} - \text{vech}\left(\mathbf{a_ia_i^T}\right) \|_2^2
\end{align}
or equivalently,  
\begin{align}
\min_{\mathbf{a_i}} \| \text{vech}^{-1}\left(\mathbf{d_i}\right) - \mathbf{a_ia_i^T} \|_F^2
\label{cov-dl-1_finalstep}
\end{align} 
The global minimum for this optimization problem is $\mathbf{\hat{a_i}} = \sqrt{\lambda_1}b_1$, where $\lambda_1$ is the largest eigenvalue of $\text{vech}^{-1}\left(\mathbf{d_i}\right)$, and $b_1$ is the associated eigenvector. For a visualization of the algorithm, see Fig. \ref{fig:cov-dl-1}.
\subsection{Undercomplete $\mathbf{D}$ (Cov-DL-2)}
\begin{figure*}[!ht]
\centering
\begin{subfigure}{.5\textwidth}
  \centering
  \includegraphics[width=.9\linewidth]{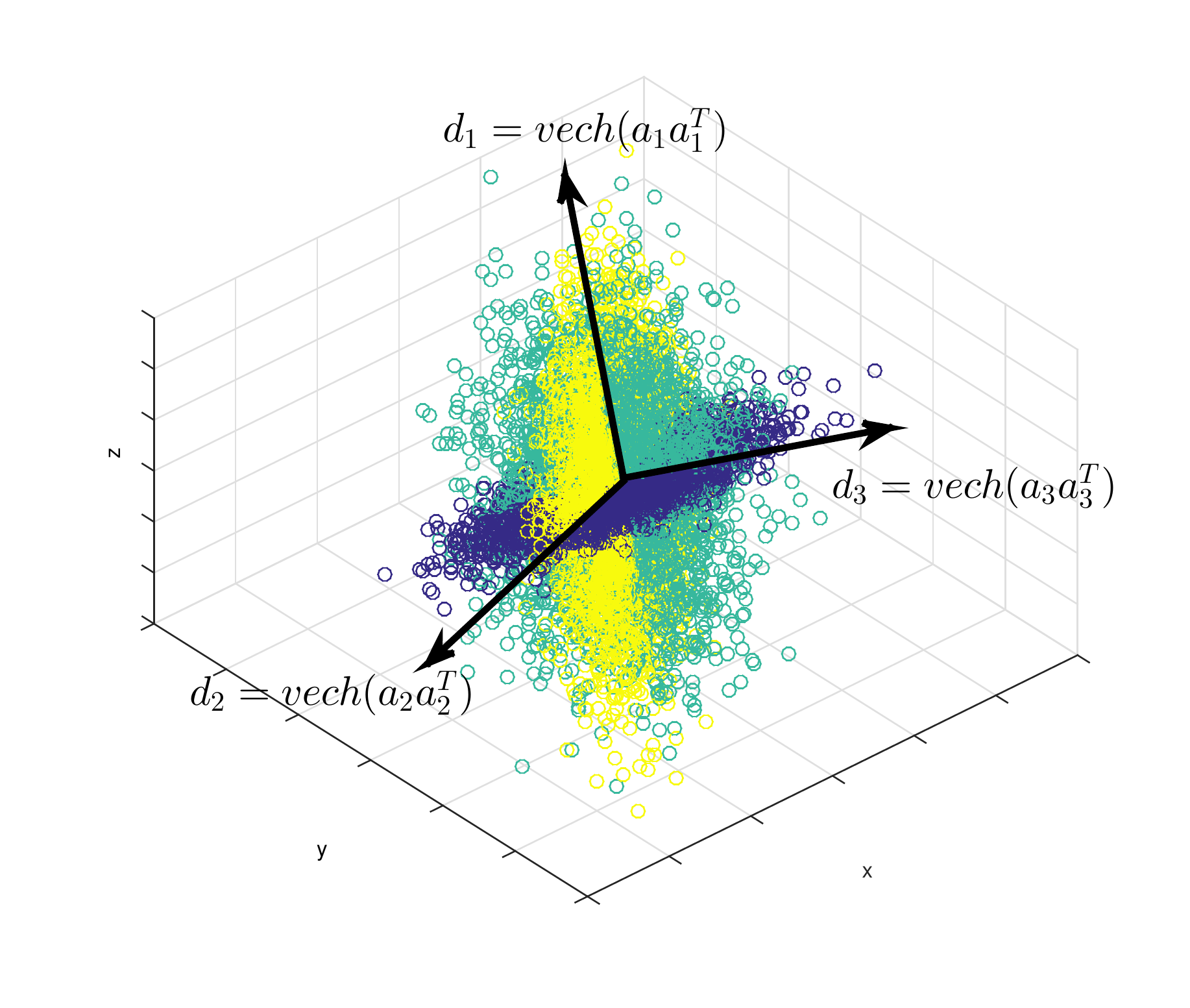}
  \caption{}
  \label{fig:cov-dl-1}
\end{subfigure}%
\begin{subfigure}{.5\textwidth}
  \centering
  \includegraphics[width=.9\linewidth]{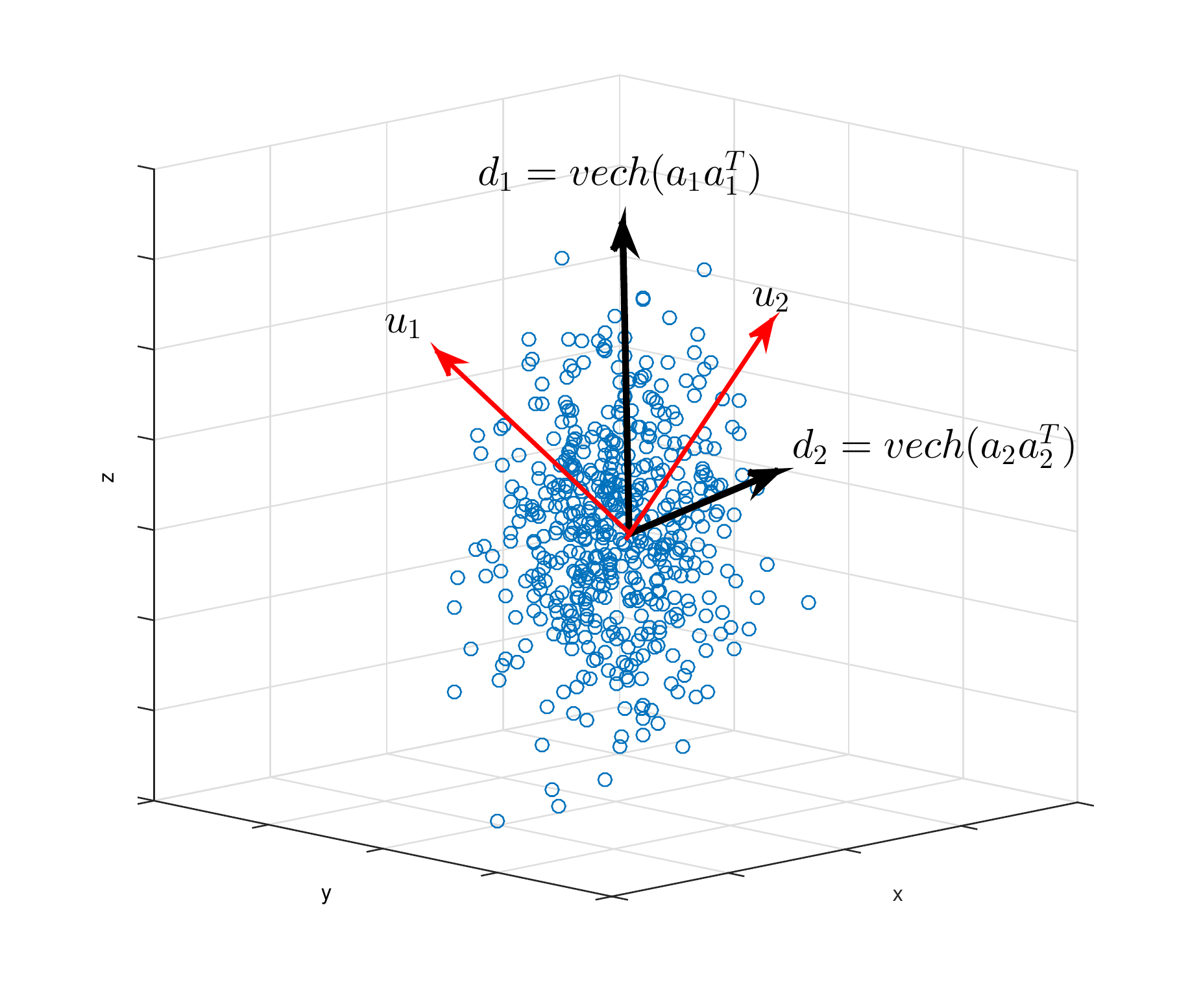}
  \caption{}
  \label{fig:cov-dl-2}
\end{subfigure}

\caption{A geometrical explanation of Cov-DL for $M = 2, k = 2$. (a) If $N = 3$, then $\mathbf{A}\in\mathbb{R}^{2\times 3}$, and $\mathbf{D}\in\mathbb{R}^{3\times 3}$. In this case $\mathbf{d_1, d_2, d_3}$ are identifiable with a dictionary learning algorithm applied on the data of vectorized outer products of segments. Associated $\mathbf{a_1, a_2, a_3}$ can then be found via solving Eg. \eqref{cov-dl-1_finalstep} (Cov-DL-1). (b) If N = 2, then $\mathbf{D}\in\mathbb{R}^{3\times 2}$, and data is not sparse since $k=N=2$. $\mathbf{D}$ is not identifiable through learning the 2-dimensional subspace (PCA results in $\mathbf{u_1, u_2}$). In this case, we solve Eq. \eqref{cov-dl-2_finalstep} to directly find $\mathbf{A}$ such that $\mathbf{D}$ will span $\mathcal{R}(\mathbf{U})$ (Cov-DL-2).}
\label{fig:test}
\end{figure*}

When, $N < M(M+1)/2$, the data points \{$\text{vech}\left(\mathbf{\Sigma_{Y_s}}\right), \forall s\}$ live on or near a subspace of dimension $N$, which is spanned by the columns of $\mathbf{D}$. We denote this subspace as $\mathcal{R}(\mathbf{D})$. We can learn $\mathcal{R}(\mathbf{D})$ with methods such as Principal Component Analysis (PCA) without imposing any sparsity constraints on $\delta_s$. However, the set of basis vectors $\mathbf{U}$ that a subspace learning algorithm, such as PCA, returns only guarantee $\mathcal{R}(\mathbf{D}) = \mathcal{R}(\mathbf{U})$, not $\mathbf{U=D}$. Therefore, we can extract $\mathcal{R}(\mathbf{D})$ but there is an ambiguity about the basis vectors $\mathbf{D}$. Note, however, that we can enforce the conditions that the columns of $\mathbf{D}$ satisfy $\mathbf{d_i} = \text{vech}(\mathbf{a_ia_i^T})$ and also span $\mathcal{R}(\mathbf{U)}$ as closely as possible. Furthermore, since the projection operator for a given subspace is unique, namely $\mathcal{R}(\mathbf{D}) = \mathcal{R}(\mathbf{U})$ if and only if $\mathbf{D(D^TD)^{-1}D^T = U(U^TU)^{-1}U^T}$, we can obtain $\mathbf{A}$ by solving the following optimization problem. 
\begin{align}
\min_{\mathbf{a_i}} &\| \mathbf{D(D^TD)^{-1}D^T - U(U^TU)^{-1}U^T}\|_F^2 \nonumber \\
&\text{s.t} \hspace{4mm} \mathbf{d_i} = \text{vech}(\mathbf{a_ia_i^T})
\label{cov-dl-2_finalstep}
\end{align}
where U is learned through a subspace learning algorithm on data points \{$\text{vech}\left(\mathbf{\Sigma_{Y_s}}\right), \forall s\}$. We compute the above cost function's gradient w.r.t $\mathbf{A}$ using the chain rule and can minimize the cost function using quasi-Newton optimization methods. We emphasize that although $\mathbf{D}$ is not overcomplete in this case, the mixing matrix $\mathbf{A}$, which relates the cortical sources to the scalp EEG sensors, can still be complete or overcomplete. For a visualization of the algorithm, see Fig. \ref{fig:cov-dl-2}.  
\subsection{Remarks}
We provide some comments about important aspects of above described algorithms. First, notice that the number of data points that Cov-DL is trained on is substantially reduced because of segmenting and learning in the covariance-domain (there is now effectively one data point per segment). For example, if $t_s$ is 4 seconds and sampling rate is 250Hz, the total number of data points used is $\frac{1}{1000}N_d$ if the segments are non-overlapping. The number of data points for Cov-DL will increase as the overlap ratio increases. However we have found that algorithm performance does not improve when the overlap ratio of consecutive segments increases beyond 0.5. The reduced number of data points in the Cov-DL-1 framework linearly speeds up the dictionary learning computation time and makes its application to EEG feasible. 

The segment length $t_s$ is an important parameter that affects the performance of the algorithms. If the segment length $t_s$ is short, the sample-covariance $\frac{1}{L_s}\mathbf{X_sX_s^T}$ is no longer pseudo-diagonal and thus the derivation in \eqref{cov-dl} is not accurate. On the other hand, as $t_s$ gets longer, the number of active sources in a segment increases (becomes less sparse), thus the performance of Cov-DL-1 will decrease. We have found that the choice $t_s \in [2, 4]$sec. provides a good compromise in our experiments.  

We also note that for both algorithms to succeed, the power of the individual sources in segments $\delta_s$ should not stay constant throughout the recording. This is required to ensure that $\mathbf{D}$ is identifiable for algorithm Cov-DL-1 and that the data points $\mathbf{\Sigma_{Y_s}}$ obtained by $\mathbf{D\delta_s}$ fill the space spanned by $\mathbf{D}$ for Cov-DL-2. This requirement holds for most EEG sources, including event-related potentials/oscillations and eye/head movement related artifact sources. To the best of our knowledge, the only EEG source that has constant power across the whole recording is electronic noise/line noise. Yet, the characteristics of this source is available (a 50Hz/60Hz sine wave) and can be filtered in the pre-processing step of EEG analysis. 

Finally we note that for algorithm Cov-DL-1, one can choose any dictionary learning algorithm for learning $\mathbf{D}$. Here, we use Bilinear Generalized Approximate Message Passing (BiGAMP-DL) \cite{parker2013bilinear}, an EM-based bayesian dictionary learning method leveraging approximate message passing. This method has the advantage of automatically learning the sparsity level and signal-to-noise ratio (SNR). For Cov-DL-2, we have used the robust PCA method described in \cite{hauberg2014grassmann} to identify $\mathbf{U}$. 

\begin{figure*}[!ht]
\centering
\begin{subfigure}{.5\textwidth}
  \centering
  \includegraphics[width=.8\linewidth]{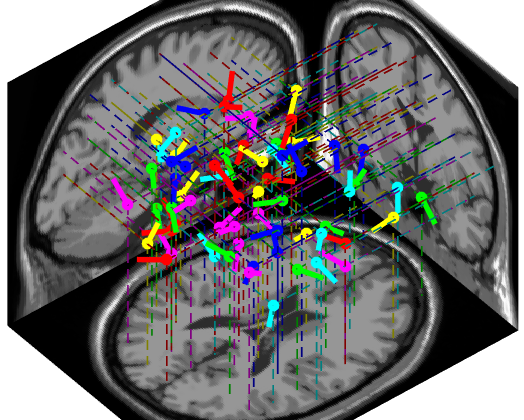}
  \caption{}
  \label{fig:simulation}
\end{subfigure}%
\begin{subfigure}{.5\textwidth}
  \centering
  \includegraphics[width=.5\linewidth]{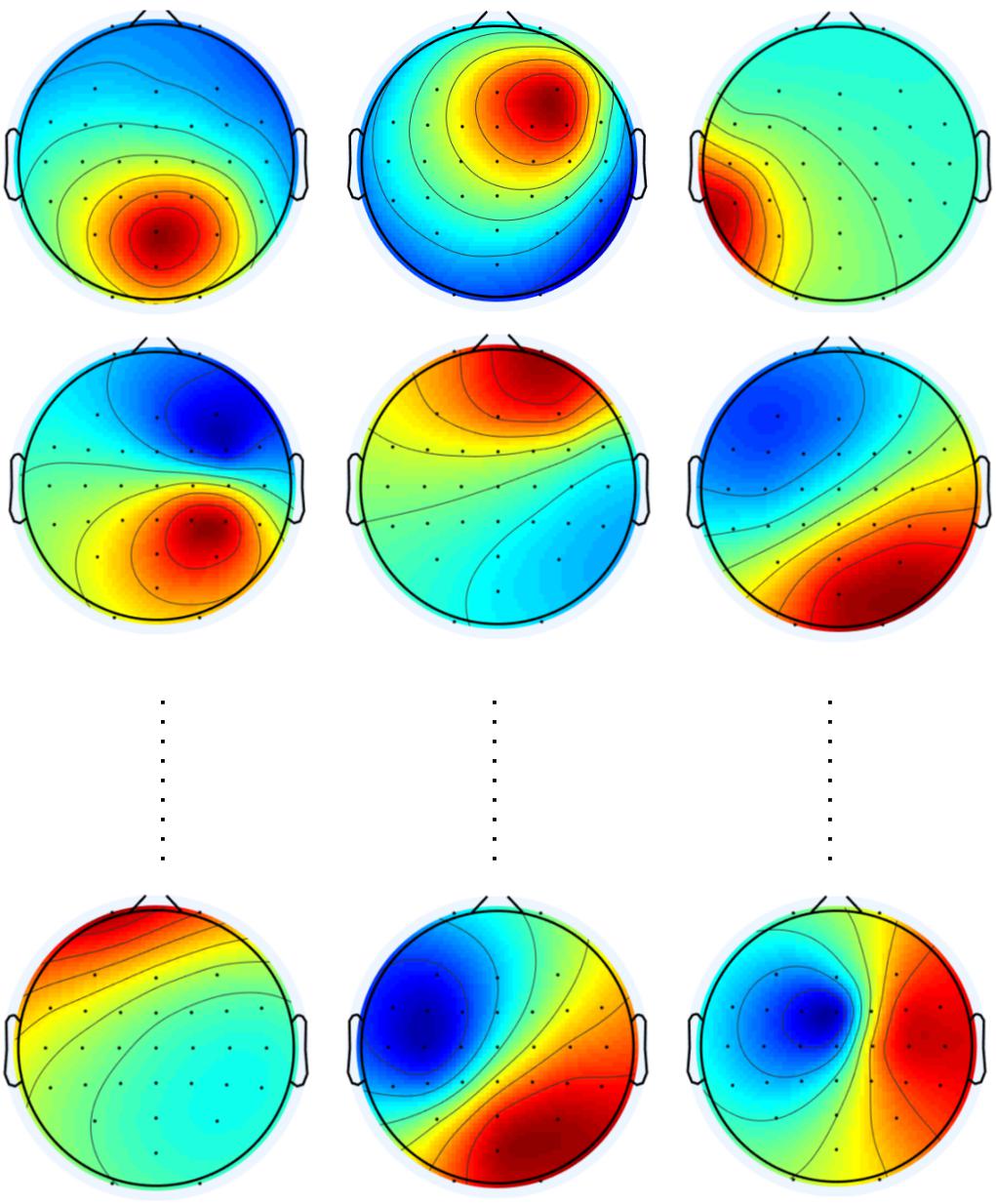}
  \caption{}
  \label{fig:scalp maps}
\end{subfigure}
\par\bigskip
\par\bigskip
\begin{subfigure}{.4\textwidth}
  \centering
  \includegraphics[width=.8\linewidth]{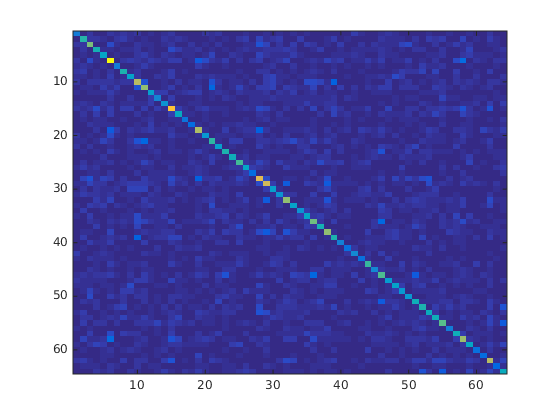}
  \caption{}
\end{subfigure} %
\begin{subfigure}{.4\textwidth}
  \centering
  \includegraphics[width=.7\linewidth]{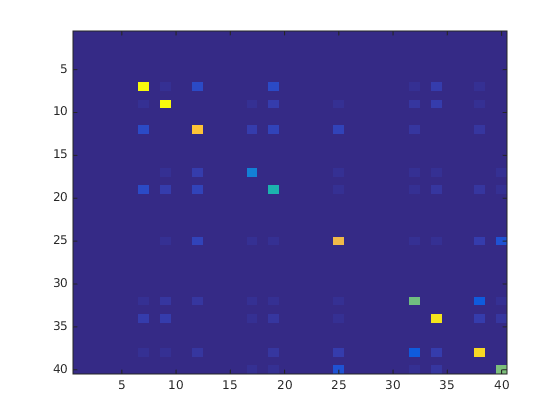}
  \caption{}
\end{subfigure} %

\caption{ (a) Randomly located and oriented $N = 64$ dipoles/sources in the MNI head model that generate the simulated EEG. (b) Some of the scalp maps (with 32 channel locations) associated with the dipoles in (a). These constitute columns of true mixing matrix $\mathbf{A}_{\text{true}}\in \mathbb{R}^{32\times 64}$. Dictionary $\mathbf{A_{\text{true}}}$ has maximum spatial coherence of 0.9888. (c) Outer product of the source matrix in a 2sec. segment (sample-covariance) from Scenario 1; $M= 32, N = 64$, $k = 64$, all the sources are active at any give time. (d) Outer product of the source matrix in a 2sec. segment (sample-covariance) from Scenario 2; $M= 8, N = 40$. $k = 10$ sources are active at any given segment.}
\label{fig:test}
\end{figure*}


\section{Experiments}
\label{sec:Experiments}
\subsection{EEG Simulation}
First we test our algorithm on three simulated data scenarios, for which we exactly know the ground truth mixing matrix $\mathbf{A}_{\text{true}}$. We simulate the placement of 32 electrodes on the scalp as shown in Fig. \ref{fig:scalp maps}. To generate the mixing matrix, we place dipolar sources in the brain using the Montreal Head Institute (MNI) head model. We assign random locations and random orientations for each dipole. Using the FieldTrip toolbox \cite{oostenveld2010fieldtrip}, we compute the projection weights of the $i$-th dipole to each channel (scalp maps) and obtain the true $\mathbf{a_i}$. See Fig. \ref{fig:scalp maps}. For realistic source activations $\mathbf{X}$, we generate an AR (auto-regressive) model via Source Information Flow Toolbox (SIFT) \cite{delorme2011eeglab} under EEGLAB \cite{eeglab} and obtain super-Gaussian source activations of duration 66 minutes with 100Hz sampling rate. We choose a segment length $t_s = 2$sec. (200 frames) and scale the sources in each segment with a random weight uniformly assigned in the continuous interval [1,2] to model the possibly varying power dynamics of brain sources across the recording.

For the first scenario, we first test and compare algorithms for the case of a complete mixing matrix ($M = N$). We select $M = N = 32$, for an overcompleteness ratio of $N/M = 1$. We also let $k = N = 32$, so that all the sources are active in any given segment. We generate scalp EEG with $\mathbf{Y = A_{\text{true}}X}$ and apply Cov-DL-2 on $\mathbf{Y}$ with $t_s = 2$sec non-overlapping segments. The accuracy of the result is measured as the ratio of the number of scalp maps that are recovered (having correlation higher than 0.99 with true scalp maps) to N. We compare our algorithm with the 1-model AMICA \cite{jasonAMICA} and RICA \cite{le2011ica}. 

For the second scenario, we have $M=32, N = 64$, and overcompleteness ratio $N/M = 2$. We also let $k = N = 64$, and again all the sources are active in any given segment. We generate scalp EEG with $\mathbf{Y = A_{\text{true}}X}$ and apply Cov-DL-2 on $\mathbf{Y}$ with $t_s = 2$sec non-overlapping segments. We compare our algorithm with the overcomplete ICA method RICA \cite{le2011ica} and concatenated dictionary obtained from multi-model AMICA ($N/M = 2$ models in this case) \cite{jasonAMICA}. 

For the third scenario, we have $M = 8, N = 40, k = 10$, and overcompleteness ratio $N/M = 5$. In each segment a randomly selected $k$ out of $N$ sources are retained and $N-k$ sources are assigned no activation. At any given segment there are more active sources than channels ($k>M$) and the set of active sources are changing throughout the recording. We select $M=8$ channels out of the 32 channels shown in Fig. \ref{fig:scalp maps} such that we uniformly cover the whole head. In this case, since $N\geq M(M+1)/2$ and $k<M(M+1)/2$, we use Cov-DL-1. We compare with the results obtained from RICA and 5-model AMICA ($N/M = 5$). 

\begin{figure}
\centering
\includegraphics[width = 0.9\linewidth]{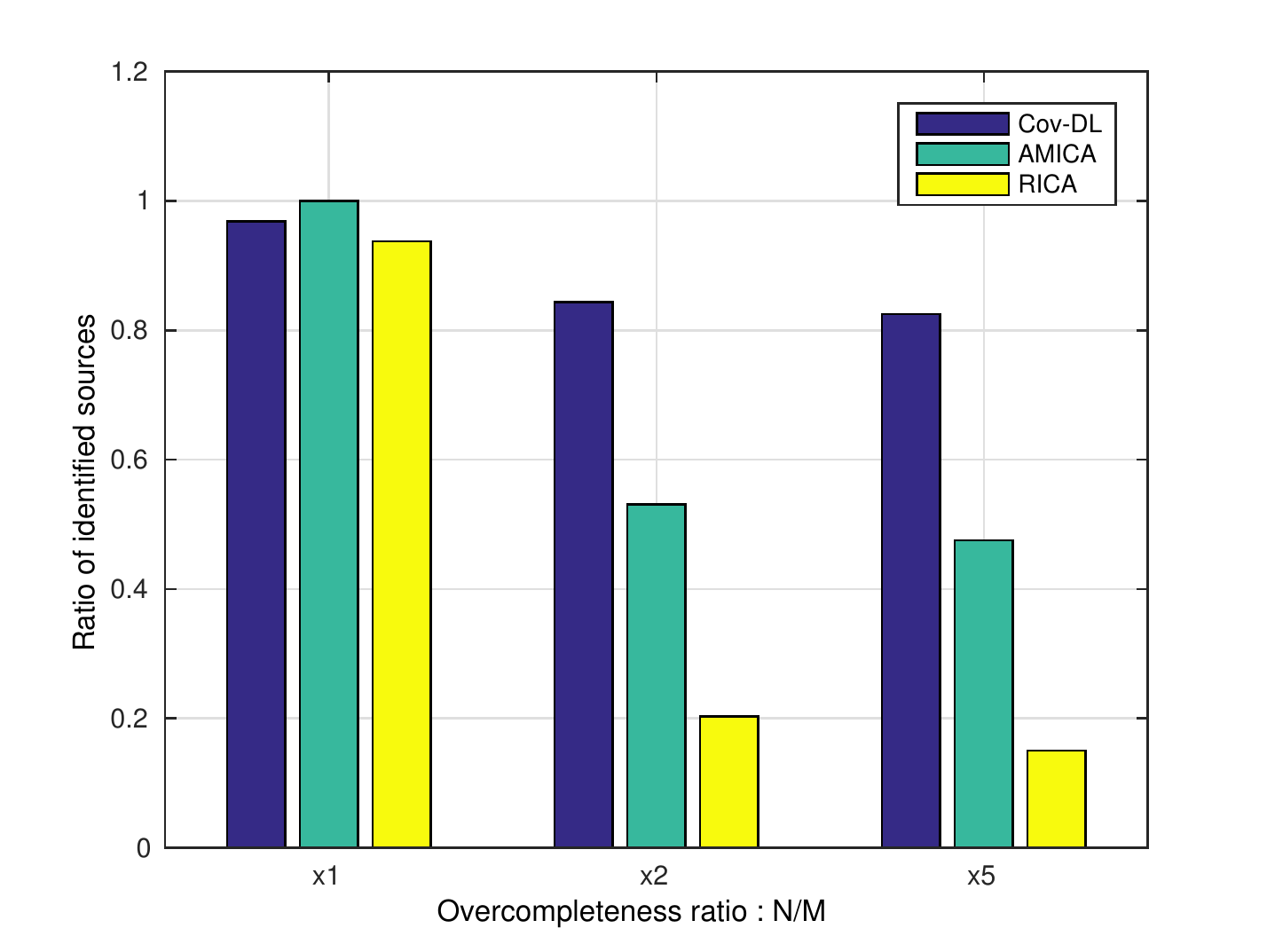}
\caption{Simulation results for three cases: complete, two times overcomplete, five times overcomplete. In the complete case. $M=32,N=32,k=32$, Cov-DL-2 is used. In the second scenario $M=32,N=64,k=64$, Cov-DL-2 is used. Third case: $M=8,N=40,k = 10$, Cov-DL-1 is used.}
\label{fig:simulationResults}
\end{figure}

The results of three scenarios are shown in Fig. \ref{fig:simulationResults}. It can be seen that when $M=N$, single model AMICA shows perfect source identification whereas Cov-DL performs slightly worse but still has accuracy of 0.9687 (recovers $31/32$ components). This might be because fewer number of data points (number of segments) are fed to Cov-DL compared to AMICA and AMICA has the ability to model arbitrary source probability densities in an adaptive way. RICA performs the worst with an identification ratio of 0.9375 even under ideal complete conditions. This is likely due to the high coherence of the realistic mixing matrix, since other experiments showed that RICA demonstrates perfect recovery with random mixing matrices (a low coherence situation). In the overcomplete scenarios, we see that there is a drop in the performance of all algorithms. AMICA and RICA perform poorly due to their differences in modeling the overcompleteness. Multi-model AMICA considers a mixture ICA model which has only few distinct states and can handle at most $M$ sources active at a given time. RICA fits a super-Gaussian distribution to sources obtained as $\mathbf{WY}$ where $\mathbf{W}$ is a tall unmixing matrix. However, sources derived in this form cannot be truly independent simply due to the necessary linear dependence of the rows of a tall matrix. Cov-DL is free of these drawbacks of existing ICA algorithms, and can handle more sources than sensors without requiring sparsity of any form as opposed to traditional dictionary learning algorithms which prohibit their use when $k>M$.

\subsection{Experiments on Real EEG}
Unlike simulated EEG, the true mixing matrix for real EEG is not known beforehand. In order to test our algorithm's performance on real EEG data, we follow the strategy proposed below.

\begin{figure}[!ht]
\centering
\begin{subfigure}{.9\linewidth}
  \centering
  \includegraphics[width=\linewidth]{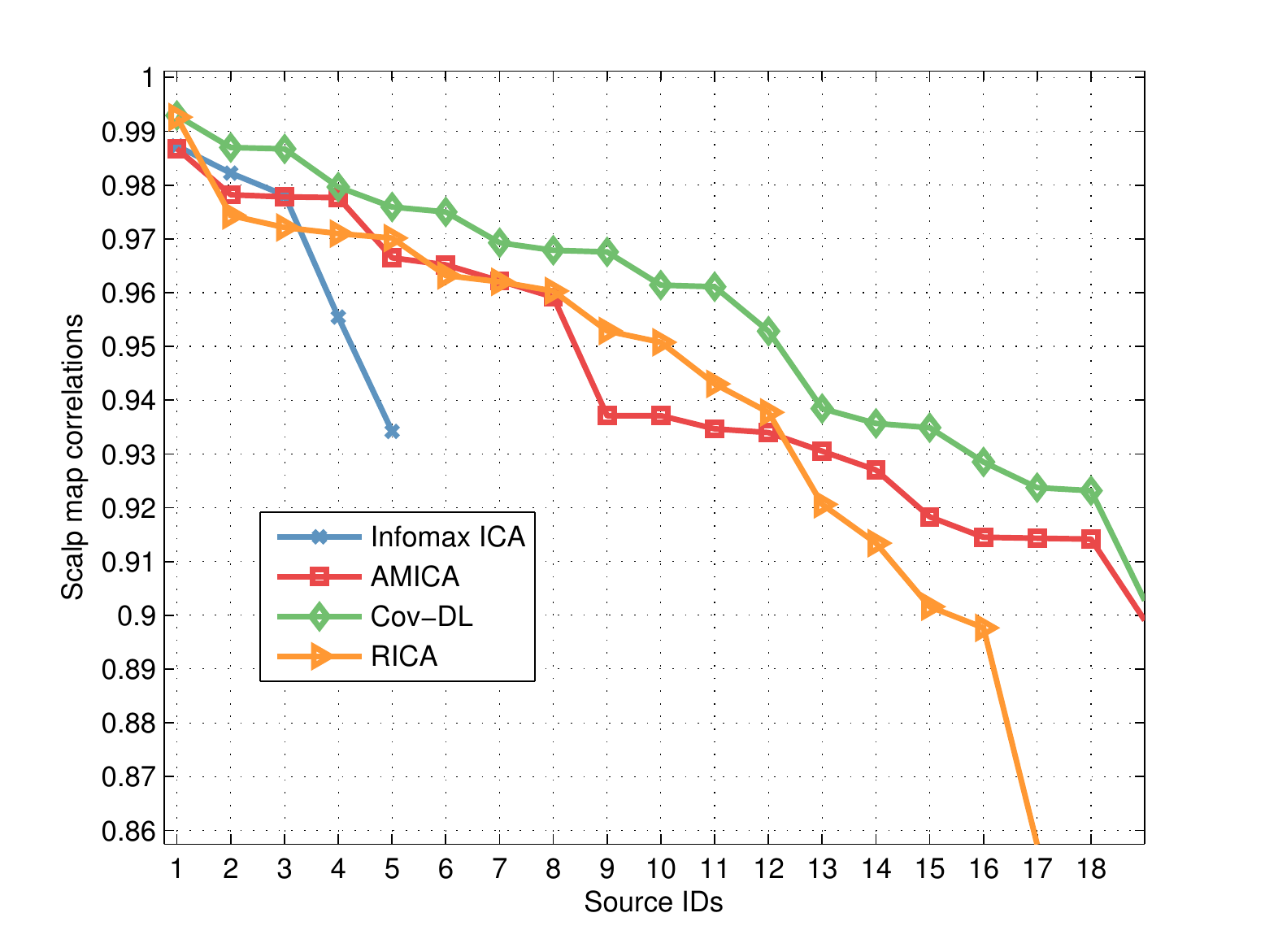}
  \caption{}
\end{subfigure}
\begin{subfigure}{.9\linewidth}
  \centering
  \includegraphics[width=\linewidth]{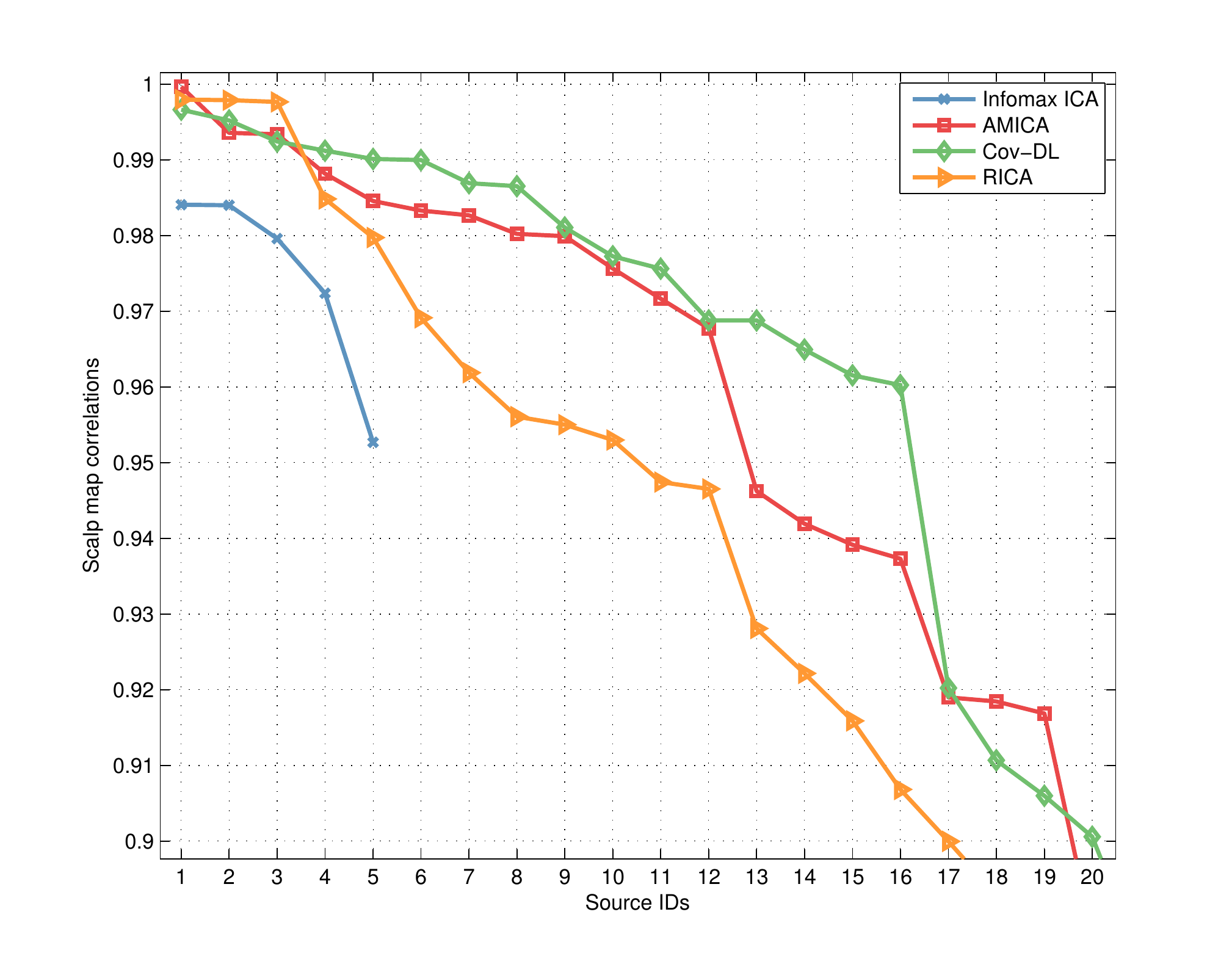}
  \caption{}
\end{subfigure}
\begin{subfigure}{.9\linewidth}
  \centering
  \includegraphics[width=\linewidth]{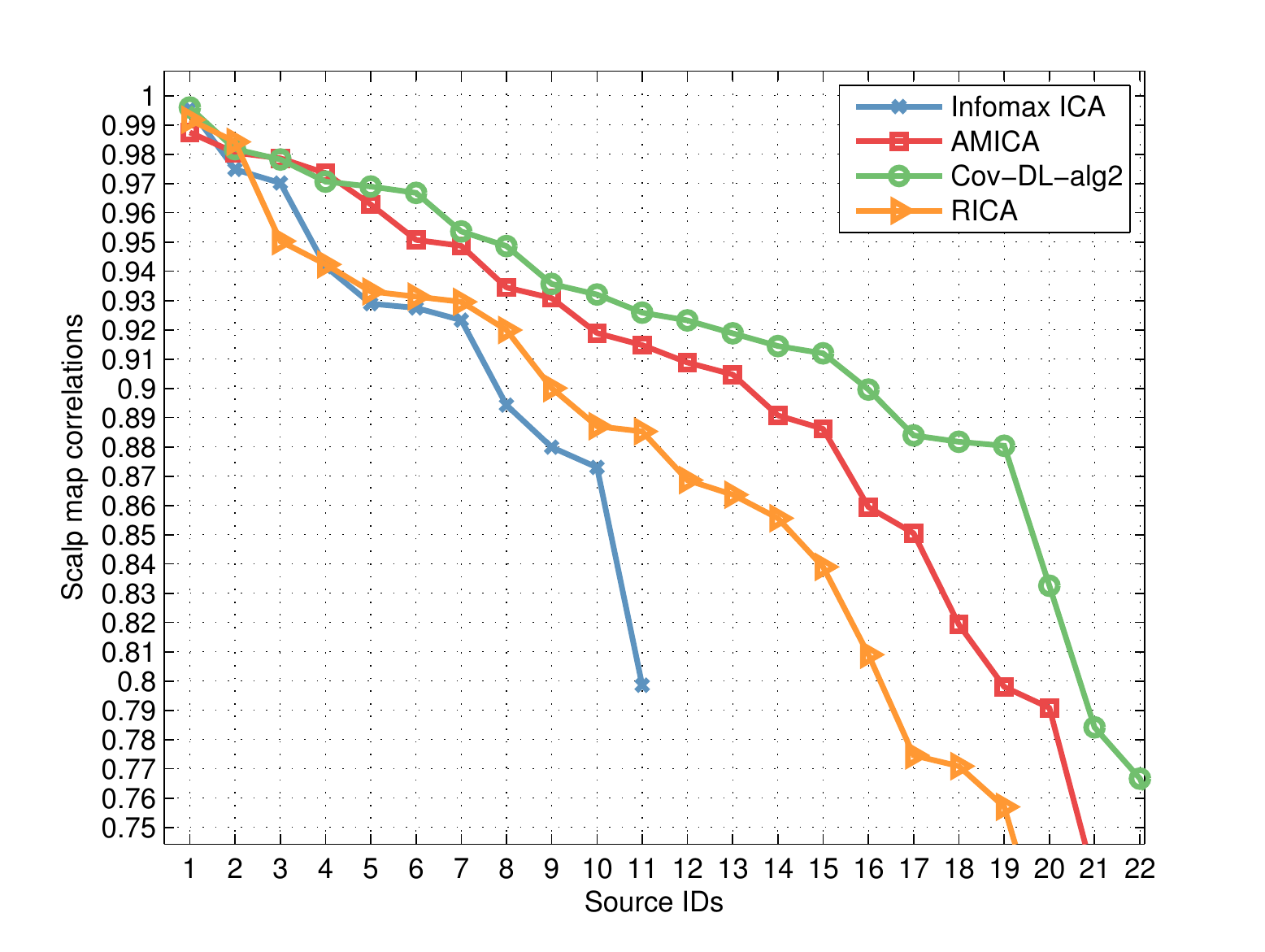}
  \caption{}
\end{subfigure}
\caption{ (a) EEGLAB sample data. $M = 5, N= 30$.  AMICA is trained with 6 models. Cov-DL-1 is performed. (b) Motor Imagery Task, $M = 5, N = 30$. Cov-DL-1 is performed. (c) Arrow Flanker task. $M = 11, N = 30$. AMICA is trained with 3 models. Cov-DL-2 is used.}
\label{fig:realData}
\end{figure}

Suppose we have an actual dataset that has $M_{\text{orig}}$ channel recordings. After rejection of the artifact windows and contaminated channels, suppose that $N$ channels remain. Then, we apply Extended Infomax ICA and extract $N$ sources and their associated scalp maps. We regard these scalp maps as ground truth mixing matrix and measure how well the proposed algorithms recover these scalp maps from using only a subset of $M$ channels out of $N$ ($M<N$). We choose $M$ channels in a spatially uniform manner as in the previous section. We compare algorithms on 3 different types of datasets; 1) EEGLAB sample data, 2) a Motor Imagery task, 3) a Arrow Flanker task. The results are shown in Fig. \ref{fig:realData}. The segment length for Cov-DL is $t_s = 2$sec, with an overlap ratio of 0.5 between consecutive segments. We plot the the sorted correlation values of resulting scalp maps with the best column match in the ground truth mixing matrix. In all 3 datasets, Cov-DL shows consistently higher correlations than multi-model AMICA and RICA. We also plot the correlation results of complete extended Infomax applied on $M$ channels to show the importance of overcomplete approaches for accurate source identification in low-density EEG systems. 

\section{Conclusion}
We proposed a dictionary learning framework, Cov-DL, that incorporates the presumed uncorrelated nature of EEG sources, which is a related but a weaker assumption than EEG source independence \cite{makeig1996independent, jasonDipolar}. Identification of the mixing matrix is carried to a higher dimensional covariance-domain, which enables source identification even if the number of sources active at any time is larger than the number of sensors - sparsity is not required. We proposed two different algorithms which depend on the relation between the number of sources targeted and number of sensors available. We have shown that the proposed algorithm Cov-DL is more successful than existing overcomplete ICA algorithms for finding the true generating matrix in EEG simulations. We have also demonstrated the power of Cov-DL on real data. The proposed algorithm, because of its ability to provide higher resolution than the number of sensors, can potentially increase the applicability of low-cost, low-density EEG systems in biomedical research.
\label{sec:Conclusion}

\ifCLASSOPTIONcaptionsoff
  \newpage
\fi



%

\bibliographystyle{IEEEbib}
\bibliography{myref}

%
%
%
%
%
%
%




\end{document}